\documentclass{article}
\pdfoutput=1

\usepackage{aimc2022}

\usepackage[utf8]{inputenc} %
\usepackage[T1]{fontenc}    %
\usepackage{hyperref}       %
\usepackage{url}            %
\usepackage{booktabs}       %
\usepackage{amsfonts}       %
\usepackage{nicefrac}       %
\usepackage{microtype}      %
\usepackage{graphicx}       %

\usepackage{graphicx}
\usepackage{subcaption}
\usepackage{mathtools}
\usepackage{multicol}
\usepackage{xspace}
\usepackage{dsfont}
\usepackage{multirow}
\usepackage{amsfonts}

\usepackage{cleveref}
\usepackage{sidecap}

\newcommand{\z}{\mathbf{z}}

\newcommand{\x}{\mathbf{x}}

\newcommand{\J}{\mathbf{J}}
\newcommand{\G}{\mathbf{G}}
\newcommand{\ourModel}{FM-Music-\!VAE\xspace}
\newcommand{\ourModelLong}{Flat-Manifold Music-\!VAE\xspace}

\DeclareMathOperator*{\argmin}{arg\,min}

\usepackage[disable]{todonotes}

\usepackage{xcolor}

\graphicspath{{figures/}}

\title{Flat Latent Manifolds for Human-machine Co-creation of Music}

\author{
\parbox{\linewidth}{\centering
Nutan Chen$^1$ 
~~~
Djalel Benbouzid$^1$
~~~
Francesco Ferroni$^2$
~~~
Mathis Nitschke$^3$
\\
Luciano Pinna$^4$
~~~
Patrick van der Smagt$^1$ 
}
\\\\
$^1$Machine Learning Research Lab, Volkswagen Group\\
$^2$ArgoAI\\
$^3$St. P\"olten University of Applied Sciences\\
$^4$St Joost School of Art and Design
}

\begin{document}

\maketitle

\begin{abstract}

The use of machine learning in artistic music generation leads to controversial discussions of the quality of art, for which objective quantification is nonsensical.
We therefore consider a music-generating algorithm as a counterpart to a human musician, in a setting where reciprocal interplay is to lead to new experiences, both for the musician and the audience.
To obtain this behaviour, we resort to the framework of recurrent Variational Auto-Encoders (VAE) and learn to generate music, seeded by a human musician.
In the learned model, we generate novel musical sequences by interpolation in latent space.
Standard VAEs however do not guarantee any form of smoothness in their latent representation.
This translates into abrupt changes in the generated music sequences.
To overcome these limitations, we regularise the decoder and endow the latent space with a flat Riemannian manifold, i.e., a manifold that is isometric to the Euclidean space.
As a result, linearly interpolating in the latent space yields realistic and smooth musical changes that fit the type of machine--musician interactions we aim for.
We provide empirical evidence for our method via a set of experiments on music datasets and we deploy our model for an interactive jam session with a professional drummer.
The live performance provides qualitative evidence that the latent representation can be intuitively interpreted and exploited by the drummer to drive the interplay.
Beyond the musical application, our approach showcases an instance of human-centred design of machine-learning models, driven by interpretability and the interaction with the end user.

\end{abstract}

\listoftodos

\section{Introduction}

Music composition is traditionally done by humans, for humans.
While tools exist which support the composer in instrumentation, counterpoint, etc., full-fledged automatic composition is somewhat beside the point, leading to existential questions on the value of art.
Still, music generation can benefit from machine augmentation, and to demonstrate this we decided to put human and machine on equal footing: can we create a jam session between two musicians, where the one is human, the other machine?

Such a machine is therefore required to generate a sequence of music, based on what's been played before. 
We address this within the framework of recurrent variational auto-encoders (recurrent VAEs) and define an optimisation scheme that results in a meaningful and interpretable latent representation. 
Our method -- we call it \ourModelLong (or \ourModel) -- infers a latent space that is isometric to the Euclidean space, hence enabling linear interpolations to be smooth and to reflect natural-sounding changes in generated musical sequences.
This work therefore takes generative sequential models (e.g., 
\citep{bayer2014learning,fraccaro2016sequential,krishnan2015deep,karl2016deep}
) a step further by not just focussing on reducing the loss of a predicted sequence, but rather quantifying the quality of prediction by the interaction between human and generative model; as it were, directly in the control loop.

Recurrent VAEs have been successful in generating long musical sequences \citep{musicVAE}, however, as for VAEs in general, they notoriously suffer from a strong regularisation \citep{2017Alemi}: the prior distribution of the latent variables, being a standard Normal distribution,
hinders the learning and prevents the posterior distribution from correctly representing the structure of the data.
As a result, interpolating in the latent space exhibits sharp and abrupt changes, which is 
unsuitable for musical applications. 

\ourModel solves the aforementioned limitation by learning a \emph{flat manifold} \citep{chen2020learning}. 
By treating the latent space as a Riemannian manifold, we constrain the topology of the decoder to be isometric to the Euclidean space.
In order to relax the over-regularisation of the standard Gaussian prior, we define the recurrent VAE as 
a hierarchical model with two stochastic layers where the prior distribution is learnt as well. 
The full model is jointly learnt with a constrained optimisation scheme.

We show that our method infers a latent representation that is used as a theoretically-motivated distance measure via a set of experiments on music data. 
Furthermore, we describe how our method was deployed in an artistic representation, with a back-and-forth `drum dialogue' between the model and a professional drummer. The drummer was not only answering to the model's musical generations but also visualise the latent space and tangibly perceive the trajectory of the interplay.

\section{Related work}

{\bf Music with VAEs}
MusicVAE \citep{musicVAE} is a method for embedding musical sequences into a latent space. It improves over sketch-RNN \citep{ha2018a} by using a hierarchical decoder allowing the generation of long musical sequence from a single point in the latent space. 
MusicVAE can also be used for interpolating between two musical sequences using its latent space, however, the VAE optimisation does not guarantee smooth changes along the latent space pathway and once can hear some abrupt changes in the generated music. Remedying this shortcoming is part of our contribution.
Noticeably, a few other works followed the line of MusicVAE in using a latent space for generating music.
GrooVAE \citep{GrooveVAE19} was proposed to humanise (or groove) drum datasets by adapting the velocity and the offset.
GrooVAE training scheme improves the disentanglement of the latent space \citep{yang2019inspecting}, however, it requires manually setting the disentanglement factors.

{\bf Interpolation in latent space}
Spherical interpolation (Slerp) was proposed for GANs and VAEs in order to improve upon the linear interpolation when generating realistic images 
\citep{white2016sampling}
It prevents sampling from the area of the latent space which is likely to generate unrealistic outputs (near the midpoint of the two selected locations \citep{white2016sampling}), which leads the output to be close to the training dataset. 
In contrast to image datasets, we expect the models to generate interesting combinations of points and have strong generalisation.
In addition, Slerp decreases the smoothness of interpolation.
Following the geodesics of VAEs was also shown to generate smooth interpolations 
\citep{chen2019fast,ChenKK2018metrics,chen2018active,arvanitidis2017latentICLR}.
These methods search for non-Euclidean trajectories as a postprocessing, i.e., once the model is trained.
The work of \citep{chen2020learning} has Jacobian/isometric regularisation. This flattens the latent space, leading to smooth outputs even with linear interpolation. Similar works have been proposed in \citep{kumar2020implicit}.
Realistic interpolation has also been improved by an adversarial regulariser \citep{2018ICAI}. These approaches improved the models during training and were able to interpolate using Euclidean trajectories after training. None of these methods have been evaluated on sequential data. Additionally, various methods of interpolation, traversal, and reconstruction in the latent space for music data were investigated, e.g., \citep{hadjeres2017glsr,engel2017latent,pati2021attribute,wang2020pianotree}.

{\bf Distance of sequences}
Dynamic time warping (DTW) \citep{muller2007dynamic} measures similarity between two temporal sequences by computing the optimal match between two sequences. Various extensions of DTW have been studied, such as triplet constraints \citep{mei2015learning}. 
Sequence distance measurement through a ground Mahalanobis metric was presented by \citet{su2019learning}. The sequence distance metrics are learnt from labelled ground truth \citep{garreau2014metric}.
Some prior work measures distances of sequences using RNNs, e.g., \citep{bayer2012learning}. In our model, we measure the sequence distances by the metric tensor of the generative model and without labels. 
In the music area, the importance of the similarity was emphasised in \citep{Anja2016}.
\citet{similarityKnees} conducts a survey of music similarity and its application to recommendation systems.

\section{Methods}

{\bf Background on VAEs}
Given the observable data $\x \in \mathbb{R}^{N_{x}}$, and the latent variables $\z \in \mathbb{R}^{N_{z}}$, we define the 
Latent-variable models
$p(\x) = \int p(\x|\z)\, p(\z)\, \mathrm{d}\z.$
Since computing $p(\x)$ is usually infeasible, the variational auto-encoders (VAEs) \citep{KingmaW14,rezende14} approximate it by a parametric inference $q_\phi(\z \vert \x)$ to maximise the evidence lower bound (ELBO):
\begin{align}
	\mathbb{E}_{p_\mathcal{D}(\mathbf{x})}\big[\log p_{\theta}(\mathbf{x})\big] \geq
	\mathop{\mathbb{E}_{p_\mathcal{D}(\mathbf{x})}}\Big[\mathbb{E}_{q_\phi(\mathbf{z}\vert\mathbf{x})}\big[\log 
	p_\theta(\mathbf{x}\vert\mathbf{z})\big]  
	\mathop{\mathbb{KL}}\big(q_\phi(\mathbf{z}\vert\mathbf{x})\|\,p(\mathbf{z})\big)\Big], 
\end{align}
where the data is $\mathcal{D} = \{\mathbf{x_i}\}^N_{i=1}$ and $p_\mathcal{D}(\mathbf{x})=\frac{1}{N}\sum_{i=1}^{N}\delta(\mathbf{x}-\mathbf{x}_i)$ represents the empirical distribution.
The approximate posterior $q_{\phi}(\z|\x)$ and the likelihood 
$p_\theta(\x|\z)$ are decoder and encoder, respectively. The distribution parameters, $\theta$ and $\phi$, are approximated by neural networks,
Usually, a Gaussian distribution is selected as the prior $p(\z)$.

\subsection{Recurrent VAEs}
\label{sec:recurrent}

Recurrent VAEs are the simplest sequential flavour of VAEs, wherein the feed-forward encoder and decoder are replaced with recurrent networks while the latent variables remain time-independent \cite{bowman2016}. 
Several architectures for the encoder and decoder are possible. 
In the present work, the decoder follows the two-stage architecture in MusicVAE \citep{musicVAE} as it is suitable for handling long sequences.
Sharing a similar architecture also helps comparing the two algorithms in the experiments.
While MusicVAE makes use of bidirectional LSTMs \citep{hochreiter1997long,schuster1997bidirectional}, we use Gated Recurrent Units (GRUs) \citep{cho2014properties} as they yield similar performance while being simpler \citep{chung2014empirical} and empirically faster to train.

The encoder receives a sequence of music data as input $\x = \{\x^1, \x^2, \dots, \x^{N_s} \}$ consisting in $N_s$ frames and $N_d$-dimensions per frame, and outputs the parameters
of the posterior distribution $q_\phi (\z \vert \x)$, the mean and standard deviation of a Normal distribution in our case. The input data is passed to stacked bidirectional GRUs, the output of which, the two state vectors, is concatenated and fed into a fully-connected layer. The last layer parametrises the mean and standard deviation of the latent variables. 

\begin{figure}[ht!]
    \centering
        \includegraphics[width=0.6\columnwidth]{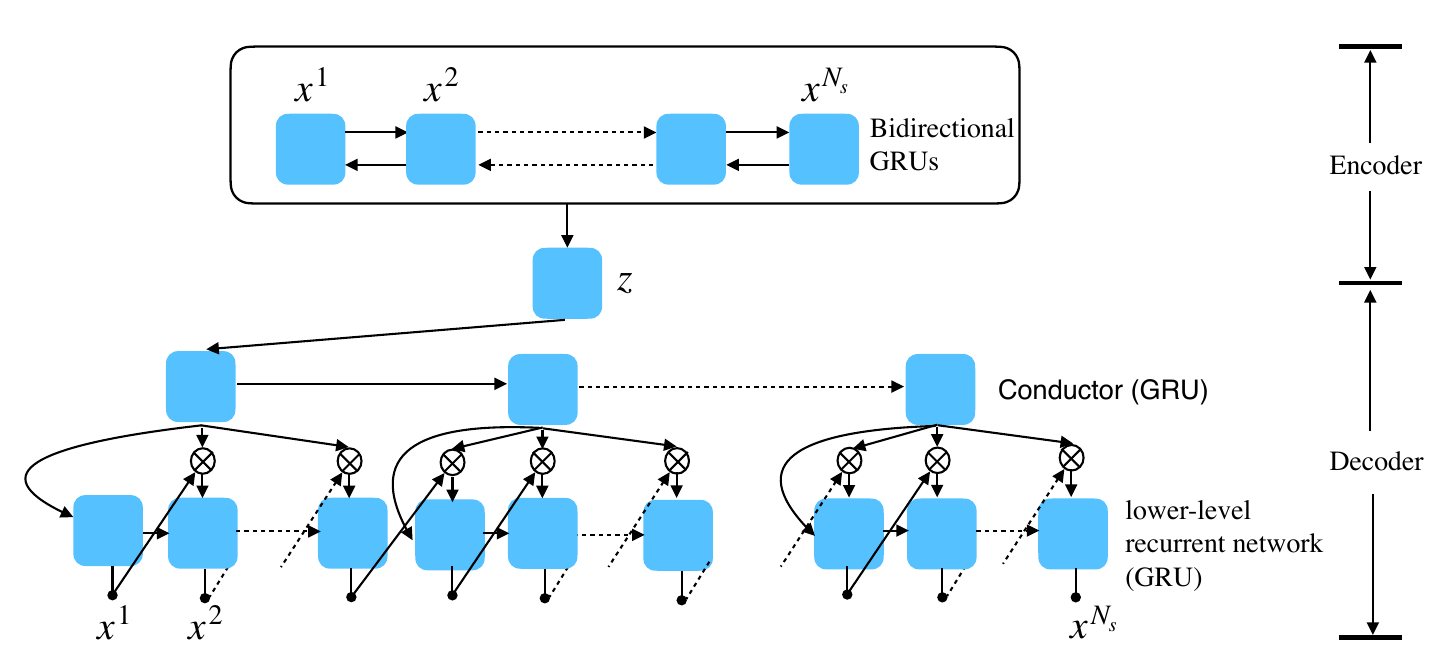}
    \caption{The architecture of \ourModel, inspired by the MusicVAE architecture. $\otimes$ denotes vector concatenation.}
     \label{fig:arch}
\end{figure}

In the decoder, a recurrent network dubbed the conductor, learns the high-level structure of the musical sequence. Every step of the conductor triggers a lower-level recurrent network which acts locally and is limited to one bar. 
Each low-level network is initialised with the hidden state of the conductor at the corresponding time and 
the conductor itself is initialised with a sample from the latent variables,
as depicted in \cref{fig:arch}.
The hierarchical decoder also prevents the so-called posterior collapse 
\citep{bowman2016,NIPS2016_ddeebdee},
i.e., when the high-capacity recurrent network of the decoder tends to ignore the input from the latent variables.

\subsection{Learning Flat Latent Manifolds with Recurrent VAEs}

\todo{intuitive explanations.}

VAEs, including their recurrent version, do not make any assumption on the inferred distances in the latent space. 
In particular, they do not guarantee that the Euclidean distance in the latent space reflects any similarity between the sequences of the observation space, hence not satisfying our initial desideratum.
Therefore, we resort to \emph{flattening} the latent manifold, i.e., constrain it to act locally as Euclidean space. 
We define the latent space inferred by the VAE as a Riemannian manifold, as motivated by \cite{ChenKK2018metrics}. Given a trajectory ${\gamma:[0, 1]\rightarrow\mathbb{R}^{N_{\z}}}$ in the latent space, the length in the observation space is
$L(\gamma) = \int_0^1 \sqrt{\dot{\gamma}(t)^T\, \G\big(\gamma(t)\big)\: \dot{\gamma}(t) } \,\mathrm{d}t$,
where $\G\in \mathbb{R}^{N_\z  \times N_\z}$ is the Riemannian metric tensor, and $\dot{\gamma}(t)$ the derivative.
The \emph{observation-space distance} is defined as the shortest possible path between two data points $D=\min_\gamma L(\gamma)$.
The trajectory ${\hat{\gamma}=\argmin_\gamma L(\gamma)}$ is defined as the (minimising) geodesic.
$\gamma$ is projected to the observation space by the decoder of VAEs. 
With the Jacobian of the decoder with respect to $\z$, the metric tensor is $\G(\z)=\J(\z)^{T}\J(\z)$.

To measure the \emph{observation-space distance} using simple Euclidean distance in latent space, we have
$D \propto \, \| \z(1)-\z(0) \|_2$,
which indicates that the Riemannian metric tensor is $\G\propto\mathds{1}$.
A manifold with this property is defined as \emph{flat manifold} \citep{lee2006riemannian}.
To learn a model with a \emph{flat latent manifold}, \citet{chen2020learning} presents
(i)~a flexible prior which allows the model to learn complex latent representations of the data by empirical Bayes;
(ii)~a regularisation $\G\propto\mathds{1}$; and (iii)~data augmentation which enables the model to have the same property in the low data density area.

\subsubsection{Hierarchical prior and constrained optimisation}
\label{ssubsec:vhp}
It has been shown that the approximated posterior can be over-regularised when the prior distribution is a standard Normal, which in turn makes the latent representation less informative \citep{TomczakW18}. 
For instance, a large distance is required for the posteriors between two data points with a large gap, if there is a lack of training data between them. In this case, the model is difficult to be trained by a standard Normal prior.
Hence, the motivation to introduce flexible priors arises, such as VampPrior \citep{TomczakW18} and VHP \citep{vahiprior2019}.

Due to its empirical results and its stable learning, we resort here to the hierarchical prior defined in \citep{vahiprior2019}, $p_\Theta(\mathbf{z}) = \int\! p_\Theta(\mathbf{z}\vert \mathbf{\zeta})\, p(\mathbf{\zeta})\, \mathrm{d}\mathbf{\zeta}$, where $p(\mathbf{\zeta})$ is the standard normal distribution.
Sampling from $q_{\phi}(\z|\x)$, we approximate the integral by an importance-weighted (IW) bound \cite{BurdaGS15}.
The upper bound on the KL term corresponding to our two stochastic layers is then, 
\begin{align}
	\mathbb{E}&_{p_\mathcal{D}(\mathbf{x})}\mathop{\mathbb{KL}}\big(q_\phi(\mathbf{z}\vert\mathbf{x})\|\,p(\mathbf{z})\big)
	\leq 
	\mathcal{F}(\phi, \Theta, \Phi)
	\nonumber
	\\
	&\equiv
	\mathbb{E}_{p_\mathcal{D}(\mathbf{x})} \mathop{\mathbb{E}_{q_\phi(\mathbf{z}|\mathbf{x})}}\bigg[
	\log q_\phi(\mathbf{z}\vert\mathbf{x}) 
	 -\mathop{\mathbb{E}_{\mathbf{\zeta}_{1:K}\sim q_\Phi(\mathbf{\zeta}|\mathbf{z})}}\Big[\log\frac{1}{K}\sum_{i=1}^{K}\frac{p_\Theta(\mathbf{z}\vert\mathbf{\zeta}_i)\, p(\mathbf{\zeta}_i)}{q_\Phi(\mathbf{\zeta}_i\vert\mathbf{z})}\Big]\bigg], 
\end{align}
where $K$ is the number of importance samples.
Based on \citep{rezende2018taming}, the ELBO is rewritten as the Lagrangian of a constrained optimisation problem:
\begin{align}
	\mathcal{L}_\text{VHP-VAE}(\theta, \phi, \Theta, \Phi; \lambda) \equiv \mathcal{F}(\phi, \Theta, \Phi) 
	+ \lambda \big( \mathop{\mathbb{E}_{p_\mathcal{D}(\mathbf{x})}} \mathbb{E}_{q_\phi(\mathbf{z}\vert\mathbf{x})}\big[\text{C}_\theta(\mathbf{x}, \mathbf{z})\big]  
	- \kappa^2 \big), 
\end{align}
with the optimisation objective $\mathcal{F}(\phi, \Theta, \Phi)$, the inequality constraint {$\mathop{\mathbb{E}_{p_\mathcal{D}(\mathbf{x})}} \mathbb{E}_{q_\phi(\mathbf{z}\vert\mathbf{x})}\big[\text{C}_\theta(\mathbf{x}, \mathbf{z})\big]\leq \kappa^2$}, and the Lagrange multiplier $\lambda$.
$\text{C}_\theta(\mathbf{x}, \mathbf{z})$ is defined as the reconstruction-error-related term in $-\log p_\theta(\mathbf{x}\vert\mathbf{z})$.
Thus, we obtain the following optimisation problem:
\begin{align}
	\min_{\Theta, \Phi} \min_{\theta} \max_{\lambda} \min_{\phi} & \: \mathcal{L}_\text{VHP-VAE}(\theta, \phi, \Theta, \Phi; \lambda)  \quad \text{s.t.} \quad \lambda \geq 0. 
\end{align}

\subsubsection{Metric tensor regularisation}

{\bf Regulariser} Having defined a constrained optimisation, it becomes straightforward to 
incorporate the aforementioned regularisation of the Jacobian into the 
learning scheme to achieve $\G\propto\mathds{1}$ \citep{chen2020learning}.
The loss function is then,
\begin{align}
	\mathcal{L}_\text{\ourModel}(\theta, \phi, \Theta, \Phi; \lambda, \eta, c^2)
	 =\mathcal{L} _\text{VHP-VAE}(\theta, \phi, \Theta, \Phi; \lambda) + \eta \, \mathcal{L}_\text{FM}(\theta; c^2)  
\end{align}
with a hyper-parameter, $\eta$, determining the influence of the regularisation, $\mathcal{L}_\text{FM}$.
\begin{align}
\label{eq:loss}
\mathcal{L}_\text{FM}(\theta; c^2) =  
\mathop{\mathbb{E}_{\mathbf{x}_{i,j} \sim p_\mathcal{D}(\mathbf{x})}} \mathbb{E}_{\z_{i,j} \sim q_\phi(\mathbf{z}\vert\mathbf{x}_{i,j})}  \bigg[  \big\|  \G\Big[sg\big(g (\z_i, \z_j )\big)\Big]  - sg(c^2)\mathds{1} \big\|_2^2 \bigg],
\end{align}
where $sg$ stands for stop gradient,
and $g$ is \textit{mixup} for exploring the latent space out of the data manifold.
The Jacobian is approximated by a stochastic method to improve the computational efficiency.
The scaling factor, $c$, is  obtained at each batch:
\begin{align}
	c^2=\frac{1}{N_\z}\,\mathop{\mathbb{E}_{\mathbf{x}_{i, j} \sim p_\mathcal{D}(\mathbf{x})}} \mathbb{E}_{\z_{i,j} \sim q_\phi(\mathbf{z}\vert\mathbf{x}_{i,j})}\Big[\mathrm{tr}\big[\G\big(g(\z_i, \z_j)\big)\big]\Big].  
\end{align}

{\bf Latent space \textit{mixup}}
The Jacobian regularisation only affects the vicinity of  the training examples, therefore, we use \textit{mixup} to augment the training data \citep{zhang2018mixup}. While the original data-augmentation method is applied to the observed data, some previous work \citep{ManifoldMixup,chen2020learning} extended it to the latent representations.
We augment the data by randomly interpolating between two points~$\z_i$~and~$\z_j$ in the latent space: 
$g(\z_i, \z_j) = (1-\alpha)\,\z_i + \alpha \,\z_j$,
with $\alpha \sim U(-\alpha_0, 1+\alpha_0)$. To compute more efficiently, we obtain $\z_i$ and $\z_j$ by shuffling a batch of $\z$.
In order to take into account the outer edge of the data manifold, we have $\alpha_0 > 0$. 
With the metric tensor regularisation in the entire latent space enclosed by the data manifold, the decoder $f(\z)$ is distance-preserving.
Given the distance in the observation space $D_\x$, and in the latent space  $D_\z$, we obtain $D_\x(f(\z_i), f(\z_j))\approx c \,D_\z(\z_i,\z_j)$.%

{\bf Jacobian approximation}
To improve the computational efficiency, we approximate the Jacobian with a first-order
Taylor expansion \citep{rifai2011higher,chen2020learning}. 
Alternatively, we use a different sampling-based approximation that scales gracefully
\citep{nesterov2017random},
\begin{align}
\J(\z) 
=\frac{1}{\mu} E_\mathbf{u}\big(f(\z+\mu \mathbf{u})\mathbf{u}\big)
\equiv
\frac{1}{\mu} E_\mathbf{u}\big([f(\z+\mu \mathbf{u})-f(\z)]\mathbf{u}\big),
\label{eq:jac2}
\end{align}
where $\mathbf{u}\in \mathbb{R}^{N_\z}$ is a normally distributed Gaussian vector and $\mu$ is a bound. 
The computation of \cref{eq:jac2} depends on the sample size used to approximate the expectation of $\mathbf{u}$. 
Therefore, for small $N_\z$, the first-order
Taylor expansion is employed, while we use \cref{eq:jac2} for large $N_\z$.

\section{Experiments and live deployment}

To assess the quality of our latent representation, we measure the smoothness and quality of the interpolation and explore how musical attributes are projected into the latent space. 
Throughout the experiments, we obtain the hyper parameters of the models
via random search.
Generated audio samples are available online\footnote{\url{https://sites.google.com/view/fm-music-vae}}.
We compare our approach with MusicVAE \citep{musicVAE}, as the SOTA model for interpolating music data.%
Quantitative evaluation is done on three music datasets.

{\bf Drum dataset}
We evaluate our model on a drum/groove dataset \citep{GrooveVAE19} with two variations  consisting in two- and four-bar sequences respectively. Each bar consists of 16 time steps. It is recorded on a Roland TD-$11^2$ electronic drum kit and we map the kit to nine categories.
In each time step, the data (MIDI notes) consists of binary-valued hits (the drum score for nine instruments), discrete-valued velocity (how strong the drums are struck), and continuous-valued offsets (indicating the micro-timing of how soon/late the note is played). Since the splits of the training, validation and testing in \citep{GrooveVAE19} easily cause overfitting and early stopping, we randomly re-split the training and validation.

{\bf Piano and cello datasets}
We obtain the piano dataset from Ryan's Mammoth of the music21 toolkit \citep{cuthbert2010music21} and the cello dataset from the Bach Cello Suites (\url{www.kunstderfuge.com}). Each dataset is augmented by pitch shifting. Therefore, the note distribution covers the full range of the 88 piano keys.
One sequence contains 4 bars, i.e., 64 time steps in total and each time step has 90 dimensions consisting of 88 `note-on', one `note-off', and one `rest' \citep{musicVAE}. 
For each dataset, we randomly split the data into 80\%, 10\% and 10\%, respectively for training, validation and testing in \cref{sec:smoothness,attribute_v}. The dataset splitting is different for interpolation quality evaluation in \cref{sec:interp_quality}.

\subsection{Evaluating the interpolations' smoothness}
\label{sec:smoothness}

\begin{figure}[ht!]
	\centering
        \begin{subfigure}[b]{0.16\columnwidth}
        \centering
	  	\includegraphics[width=\textwidth]{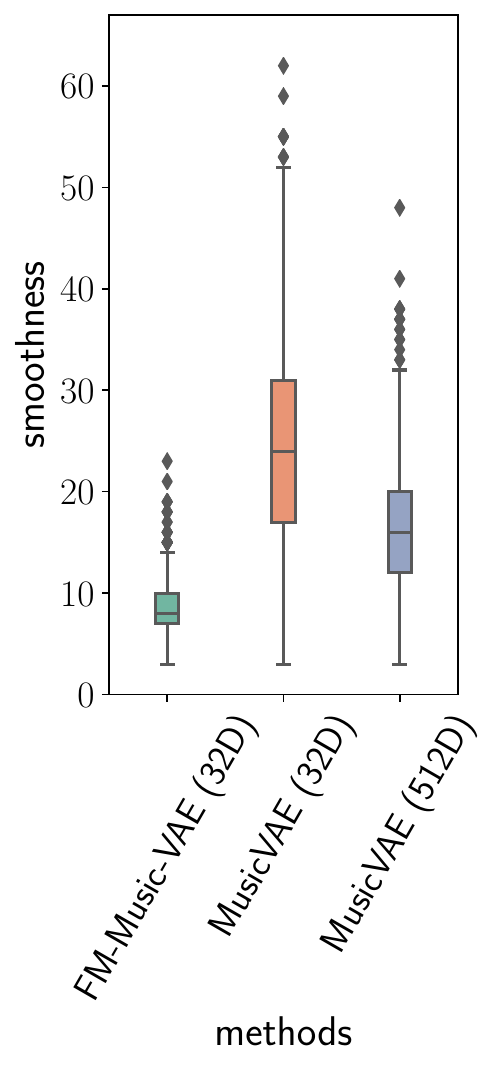}
		    \caption{}%
        \end{subfigure}
        \begin{subfigure}[b]{0.16\columnwidth}
        \centering
		\includegraphics[width=\textwidth]{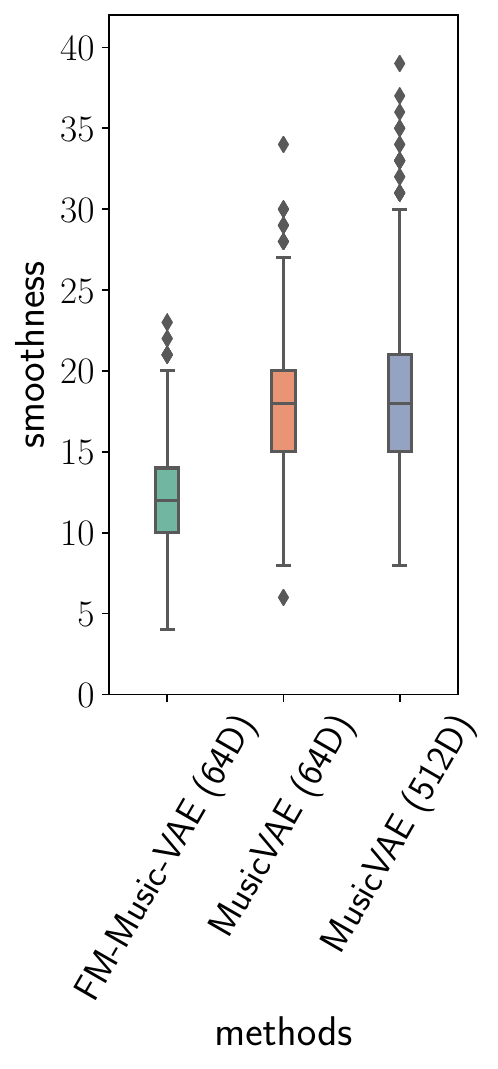}
		    \caption{}%
        \end{subfigure}
        \begin{subfigure}[b]{0.16\columnwidth}
        \centering
		\includegraphics[width=\textwidth]{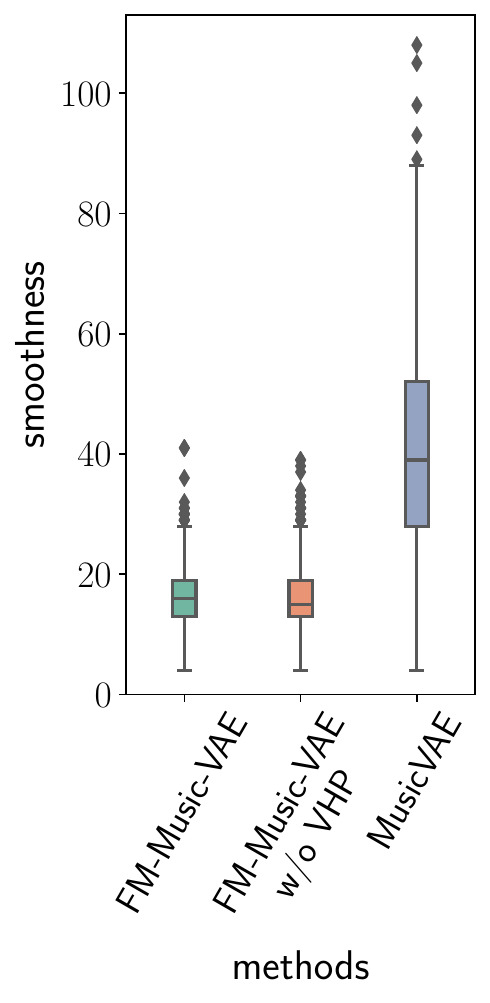}
                \vspace*{-0.2cm}
		    \caption{}%
        \end{subfigure}
        \begin{subfigure}[b]{0.16\columnwidth}
        \centering
		\includegraphics[width=\textwidth]{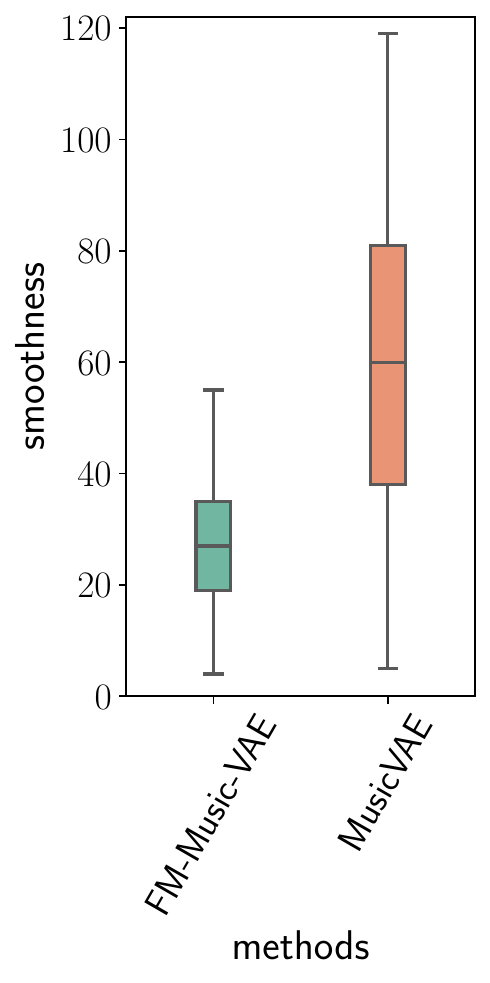}
\vspace*{-0.1cm}
         \caption{}%
        \end{subfigure}
	\caption{A comparison of the smoothness of \ourModel and MusicVAE. The brackets indicate the dimensionality of the latent space. (a) and (b) are drum and piano data, respectively, with two bars and various latent dimensions. (c) drum data with four bars and 32 latent dimensions. (d) cello data with four bars and 64 latent dimensions.}
	\label{fig:smoothness}
\end{figure}

We take the hits and on-notes for the drum and piano datasets respectively to compute smoothness.
Since they are binary data,
we define the smoothness of a sequence as 
$\max_{t\in [0, 1 \dots, N-1]}D_{\mathrm{h}}(\x_t, \x_{t+1})$,
where $D_\mathrm{h}$ represents Hamming distance, and $\x_t$ is a sequence at step $t$ of the interpolation. \Cref{fig:smoothness} shows the smoothness of 1\,000 pairs and 20 interpolated points. Interpolated samples are shown in \cref{fig:interp_samples}. With MusicVAE, the smoothness improves with a sufficiently large number of dimensions in the latent space, however, our approach is able to get overall better smoothness with a significantly lower-dimensional latent space. 
In this experiment, MusicVAE uses linear interpolation instead of the Slerp for the smoothness evaluation, since we expect to explore the whole latent space for more creativity instead of only interpolating between two points and generate data similar to the training dataset.

\subsection{Evaluating the interpolations’ quality}
\label{sec:interp_quality}

Evaluating the quality of the interpolations via human ratings is subject to a substantial bias. Hence, we quantify the quality of the interpolation by measuring some generalisation properties in between the training data points, namely, the velocity and offset for the drum data, the pitch for the piano, and the notes density for both. 
All of these properties are part of the input data except for the notes density which we manually count for each two bars.
Using the middle decile or the middle 50 percentile as a ground truth of testing datasets, we obtain the samples between the other two ends (training and validation datasets) of the distribution for each of the aforementioned properties.
The results for each property are summarised in \cref{tab:generalisation}.
Since the piano and cello datasets are one-hot encoded, the reconstruction error rate $\delta$ is reported for the whole frame, i.e., $\delta = {N_f}/{N_s}$, with $N_f$ being the number of the incorrectly reconstructed frames in a sequence. On the other hand, the error rate for the drum data is computed on a per-beat level for the density attribute, i.e., $\delta={N_{fp}}/({N_x N_s})$, where $N_{fp}$ is the number of wrongly predicted beats in a sequence. As for the velocity and offset, we report the mean square errors (MSEs) per beat. All the reported numbers, the mean and standard deviation, are evaluated over the test datasets.

\begin{table*}[ht!]
\caption{{\small Evaluating the quality of the interpolation. 
The bold font indicates the best result. The models of the drum datasets are 32D latent space, while of the piano and cello datasets are 64D latent space. The first column shows the dataset and the percentile used for test datasets.}}
\begin{center}
\begin{footnotesize}
\vskip -0.01in
\begin{sc}
{\tiny	
\begin{tabular}{l|ccccccc}
\toprule
dataset (percentile) & model 	& density	& velocity 		& offset \\
Drum, four bars & \ourModel 	&  \textbf{0.039 $\pm$ 0.023} 	&  \textbf{0.0116	$\pm$ 0.0080}	&  \textbf{0.0039 $\pm$ 0.0022} \\  
(10\%) & MusicVAE 	&   0.051 $\pm$ 0.031	&   0.0151 $\pm$ 0.0123	& 	0.0042 $\pm$ 0.0024 \\
\midrule
Drum, four bars	&  \ourModel 		& 	 \textbf{0.064 $\pm$ 0.030}& \textbf{0.0195 $\pm$ 0.0116}	&\textbf{0.0050 $\pm$ 0.0027} \\
(50\%)	& MusicVAE 		& 	0.094 $\pm$ 0.037		&	0.0286 $\pm$ 0.0163 &0.0058 $\pm$ 0.0031 \\
\midrule
dataset (percentile) & model 	& density	& pitch & \\
Piano, two bars &  \ourModel  &  \textbf{0.195 $\pm$ 0.093} &  \textbf{0.165 $\pm$ 0.099} & \\
(50\%)	& MusicVAE  & 0.234 $\pm$ 0.222 & 0.431 $\pm$ 0.088 & \\
\midrule
Cello, four bars &  \ourModel  &  \textbf{0.145 $\pm$ 0.183} &   \textbf{0.117 $\pm$ 0.165} & \\
(50\%)	& MusicVAE  & 0.616 $\pm$ 0.194 & 0.560 $\pm$ 0.211 & \\
\bottomrule
\end{tabular}
}
\label{tab:generalisation}
\end{sc}
\end{footnotesize}
\end{center}
\end{table*}

\subsection{Exploiting the structure of the latent space with attribute vectors}
\label{attribute_v}

\begin{table*}[ht!]
\caption{{\small The correlation of the attribute and the latent space (Pearson's correlation coefficient). The bold font represents the best result.}}
\begin{center}
\begin{footnotesize}
\begin{sc}
{\tiny
\begin{tabular}{l|ccccccc}
\toprule
dataset & model	& density 	& velocity		& offset	& pitch \\
Drum, two bars & \ourModel (32D)	&  \textbf{0.89} 	& \textbf{0.84}		& \textbf{0.69}& -	\\  
& MusicVAE (32D)	&  0.50 & 0.27	& 0.21	& -\\
& MusicVAE (512D)&  0.74   & 0.56		&0.52 & -\\
\midrule

Drum, four bars & \ourModel (32D)	&  \textbf{0.89} 	& 0.87		& \textbf{0.72} & -	\\  
&  \ourModel w/o VHP (32D)	&   0.87& 	\textbf{0.88}& 	0.60 & -\\
& MusicVAE (32D)	&   0.73& 	0.31& 	0.59  & -\\
\midrule
Piano, two bars & \ourModel (64D)	&  \textbf{0.89} & -  & -& \textbf{0.77} \\  
& MusicVAE (64D)	&  0.37 & -  & -	& 0.42	\\
& MusicVAE (512D)&  0.48 & -  & -	& 0.50      \\
\midrule
Cello, four bars & \ourModel (64D)	&  \textbf{0.91} &-&-& \textbf{0.81} 	\\  
& MusicVAE (64D)	&  0.79&-&-	& 0.60		 \\
\bottomrule
\end{tabular}
}
\label{tab:correlation}
\end{sc}
\end{footnotesize}
\end{center}
\end{table*}

The structure of the latent spaces can be exploited using attribute vectors \citep{musicVAE,white2016sampling}. In particular, we exploit the density, velocity and offset attributes in the drum dataset, as well as the density and pitch attributes in the piano dataset. The density attribute is computed by summing the hits (drum) or on-notes (piano) over two bars. The velocity/offset/pitch attributes are the mean over the velocities/offsets/pitches of each sequence where notes exist.

\begin{figure}
\centering
\begin{minipage}{.495\textwidth}
  \centering
    \begin{subfigure}[t]{\columnwidth}
        \includegraphics[width=\textwidth]{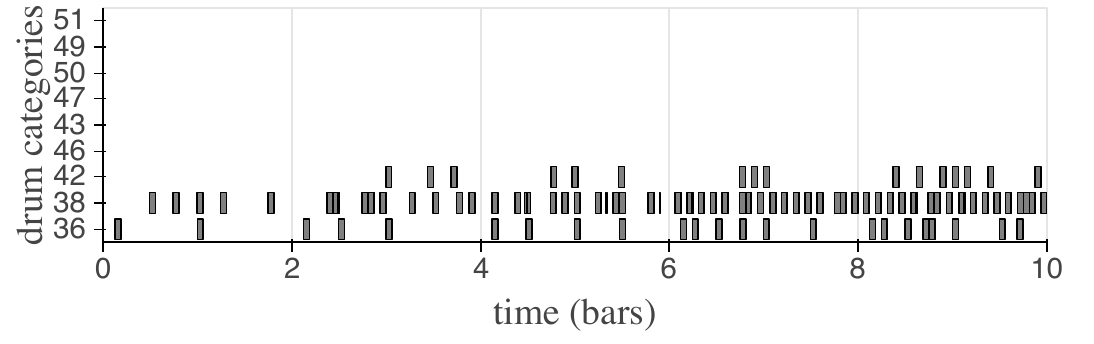}
        \caption{\ourModel (32D)}%
        \label{fig:density_musicfmvae}
    \end{subfigure}
    \\
    \begin{subfigure}[t]{\columnwidth}
        \includegraphics[width=\textwidth]{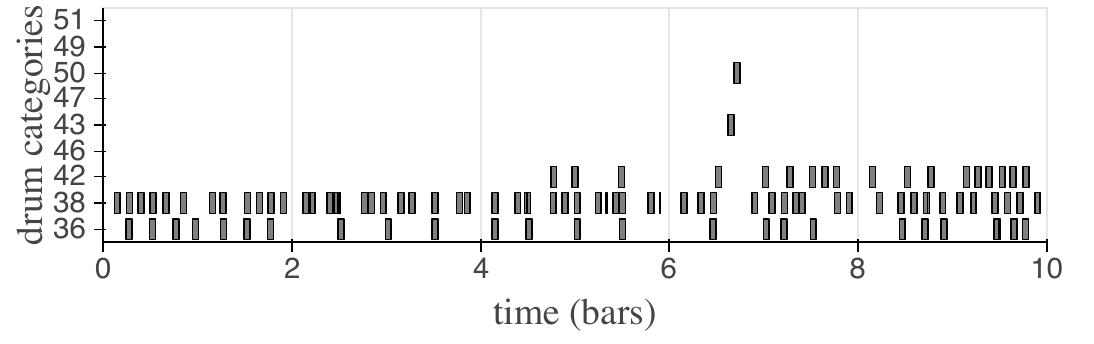}
        \caption{MusicVAE (32D)}%
        \label{fig:density_musicvae}
    \end{subfigure}
    \caption{Density vector. The middle data point (two bars) of each subfigure is from the datasets. From the left to the right directions are the attribute vectors with the scale factor from $-1$ to 1. }
     \label{fig:density_music}
\end{minipage}%
\hfill
\begin{minipage}{.48\textwidth}
    \centering
    \begin{subfigure}[t]{\columnwidth}
        \includegraphics[width=\textwidth]{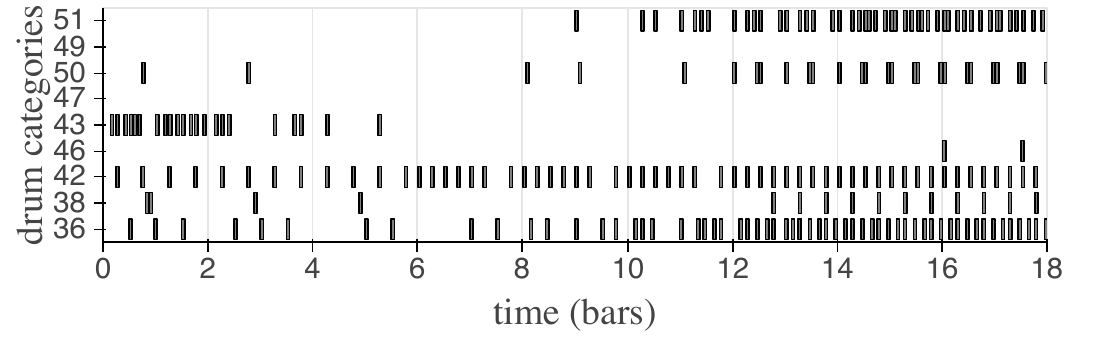}
        \caption{\ourModel (32D)}%
        \label{fig:interp_musicfmvae}
    \end{subfigure}
    \\
    \begin{subfigure}[t]{\columnwidth}
        \includegraphics[width=\textwidth]{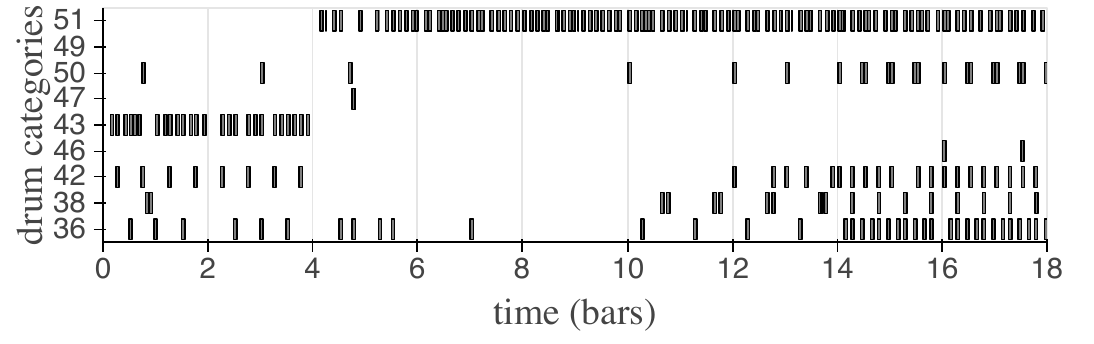}
        \caption{MusicVAE (32D)}%
        \label{fig:interp_musicvae}
    \end{subfigure}
    \caption{A comparison of music interpolation. Given the starting and ending sequences, seven points are linearly interpolated in the latent space.
    }
     \label{fig:interp_samples}
\end{minipage}
\end{figure}

5\,000 samples are selected to compute the highest and lowest 25\% percentile for each attribute, and the mean positions are computed accordingly in these regions of the latent spaces. The direction is then calculated based on the centroids.
Based on a data point in the latent space, previous work \citep{musicVAE} subtracts or adds a vector to evaluate the features. Motivated by the continuous nature of music (in contrast to discrete image features such as wearing or not wearing glasses), we instead randomly scale the vector length with a scale factor in the range of $[-1, 1]$, and then compute the linear correlation between the vector length in the latent space and attributes in the observation space. If the data point with the scaled attribute vector in the latent space exceeds the range of any axis, the corresponding sample is discarded. 
In the end, 5\,000 samples are used to calculate the correlation. 
\Cref{tab:correlation} indicates that the latent space of the proposed approach is more structured. The 2-tailed p-values are zero or close to zero, so they are not shown in the tables. \Cref{fig:density_music} illustrates an example of moving in the latent space along the density vector. The \ourModel retains the patterns of the illustrated data, while the density changes.

{\bf Ablation study}
Motivating our design choices, we conduct an ablation study in order to assess the effect of the hierarchical prior described in \cref{ssubsec:vhp}. 
\Cref{tab:correlation} shows that while \ourModel with a standard Normal prior does improve over the state-of-the-art, adding the hierarchical prior further improves the performance. Hence, the rest of the experiments are conducted with the full model.

\subsection{Drummer-model interactive deployment}

\begin{figure}[ht!]
    \centering
        \includegraphics[width=0.7\textwidth]{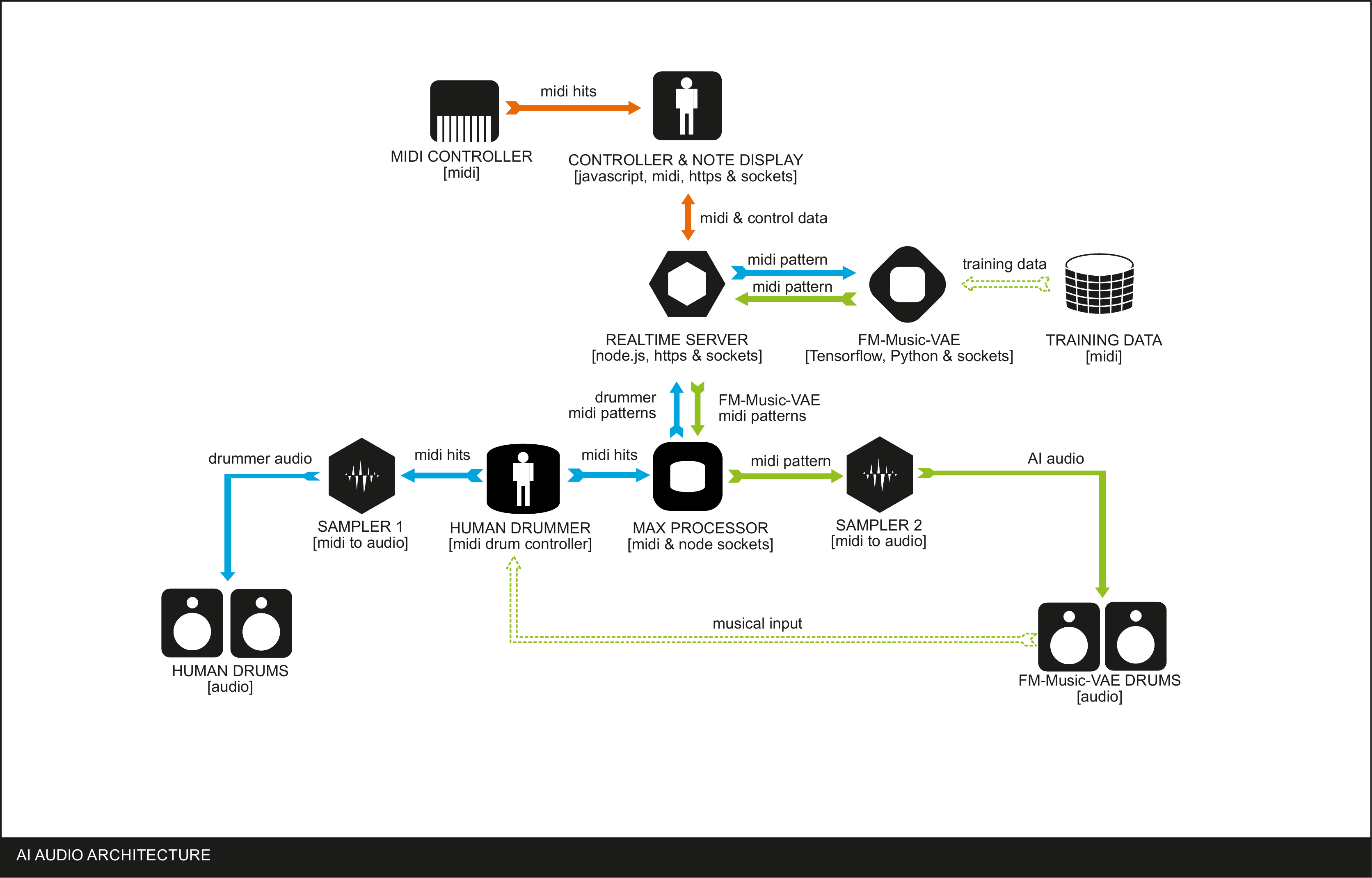}
    \caption{The architecture of the jam session.}
    \label{fig:jamsession}
\end{figure}

The final test of \ourModel was in a live artistic event. During the session,
a professional drummer interacted with our model by playing two bars of drum music, to which our model `responded' with another generated two-bars drum sequence. 
Once received, the drummer's playing was mapped into a point in the latent space. Using this point as the centre of a Gaussian distribution, a second point was then sampled and decoded into a drum sequence that the drummer could listen to -- and hopefully be inspired by.
The back-and-forth interplay lasted for about ten minutes. 
While playing, the drummer was also able to visualise the three-dimensional embedding of his own playing as well as that of the model's generations. The set of embedded points were also visually linked to reflect their occurrence in time, hence forming the two `trajectories' of the two players in a three-dimensional space. The goal was to let the drummer try to `influence' the direction the model was taking, a task that is only feasible if the latent representations can be properly interpreted.

To facilitate the musical dialog between a human drummer and the \ourModel, the infrastructure was built up around realtime socket connections. The latency, introduced by processing of the data, was overcome by working in bars of notes instead of sending over individual hits. The drummer played on a midi drum set and the individual notes were captured by a Max application and sequenced into two bars. The bars were sent over to a central server which acted as a broker between the AI model and the human drummer. The forwarded bars were processed by the model into two response-bars and sent back to the Max application in reverse route. For both human drummer and AI a separate sampler was used to convert the midi into hearable audio.
\Cref{fig:jamsession} depicts the architecture deployed during the live session.

Although the performance itself excludes any scientific pretension, we argue that the deployment of our model in this context brings value on three different fronts. First, it shows that our latent-variable model can be used for interpretability purposes, even by a non-machine-learning practitioner, as it exhibits meaningful distances. We also make the point that deploying models is valuable in itself, even in the context of scientific research, as it is anchored in real needs \citep{wagstaff2012machine}.
Finally, contributing to a novel musical experience in which an artist has the possibility to explore music both audibly and visually constitutes an artistic value.
An excerpt of the performance video is available online.

\section{Conclusion and future work}

We presented \ourModel, a theoretically-motivated approach to live interaction with a human musician while learning a high-level music representation. Our method in effect learns 
a distance metric for sequence data that allows for smooth and meaningful linear interpolations. 
Using drums and piano datasets, we experimentally demonstrated that our principled constrained optimisation scheme properly regularises the decoder and thus yields realistic music sequences.
Finally, the interactive music session between the model and a drummer illustrates that the proposed approach is able to be intuitively used by a human, although a more thorough and scientific study is needed to support this initial observation.
In the future, using high-level control in latent space, e.g.\ through, reinforcement learning, for generating longer musical sequences would be a natural follow-up.

\todo{While future work is discussed, the paper lacks a limitations section. The limitations are hinted at in diff erentparts of the paper, but there is no specifi c section that discusses them in length.}

\section*{Acknowledgements}
The authors would like to thank Matthias Lachenmayer who is the professional drummer for testing our model. We thank Alexej Klushyn for some code of the related work. We thank Botond Cseke for useful discussions.

\bibliographystyle{named}
\bibliography{recurrentfmvae}

\appendix
\newpage
\cleardoublepage
\appendix
\onecolumn

\section{Appendix}

\subsection{Interpolation of the drum dataset}
\label{app:interp_drum}

\begin{figure}[ht!]
    \centering
    \begin{subfigure}[b]{0.48\columnwidth}
        \includegraphics[width=\textwidth]{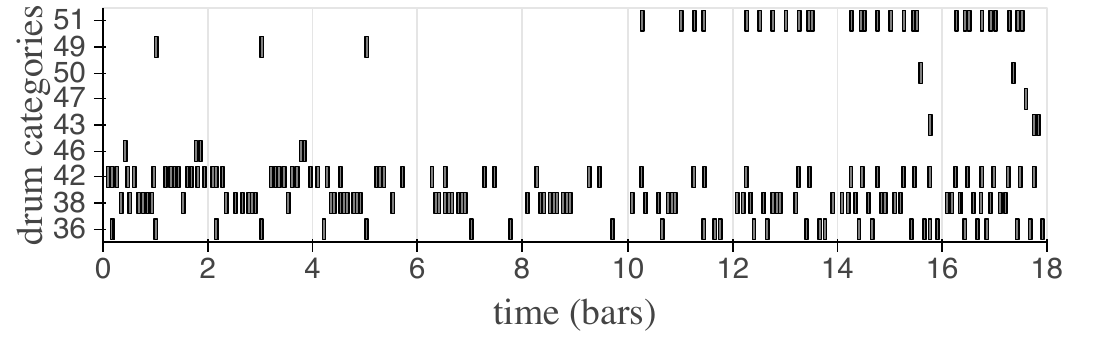}
        \caption{\ourModel}%
    \end{subfigure}
    \\
    \begin{subfigure}[b]{0.48\columnwidth}
        \includegraphics[width=\textwidth]{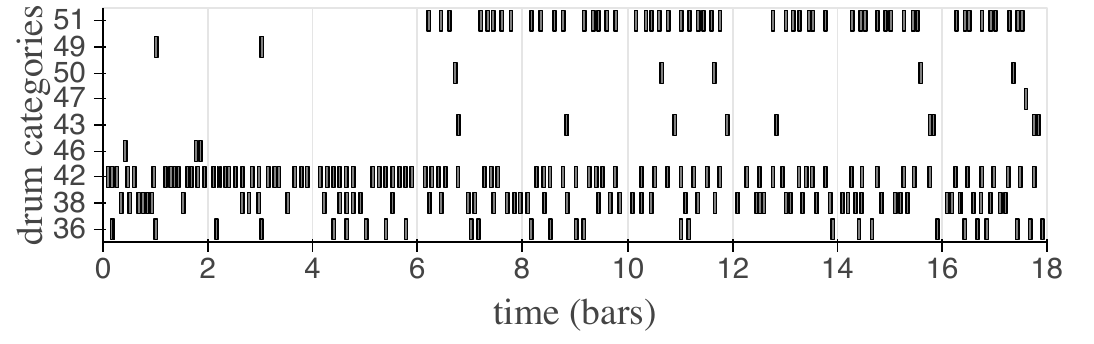}
        \caption{MusicVAE (32D)}%
    \end{subfigure}
    \\
    \begin{subfigure}[b]{0.48\columnwidth}
        \includegraphics[width=\textwidth]{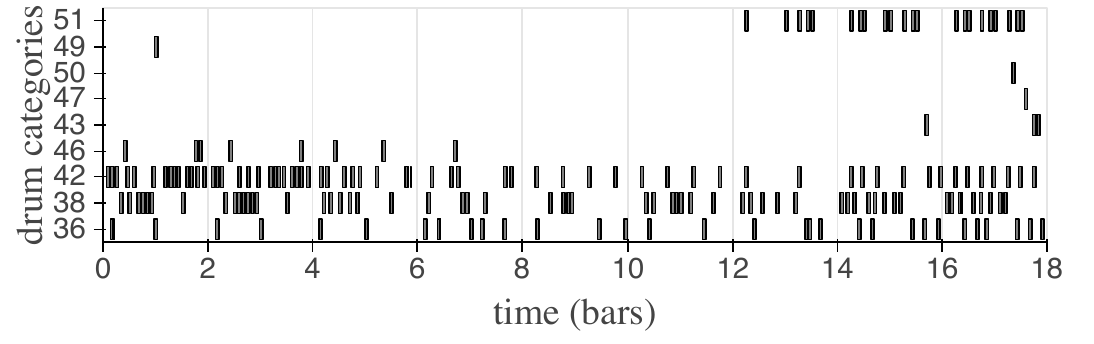}
        \caption{MusicVAE (512D)}%
    \end{subfigure}
    \caption{Drum interpolation. See \cref{fig:interp_samples} for more details.}
\end{figure}
\begin{figure}[ht!]
    \centering
    \begin{subfigure}[b]{0.48\columnwidth}
        \includegraphics[width=\textwidth]{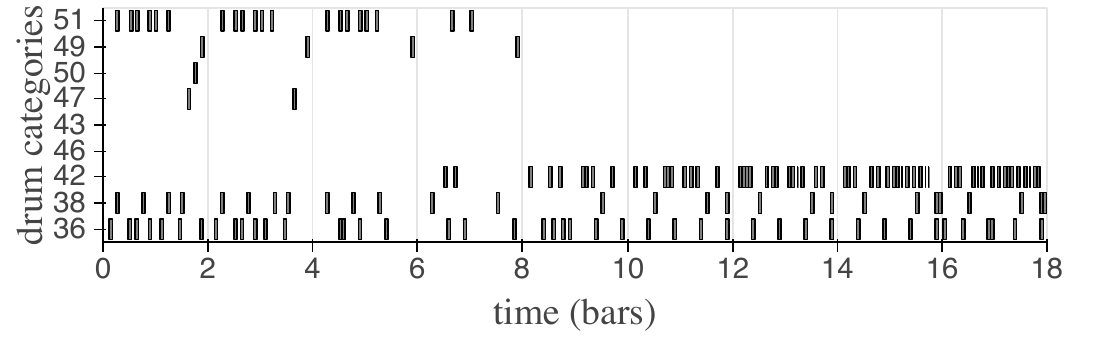}
        \caption{\ourModel}%
    \end{subfigure}
    \\
    \begin{subfigure}[b]{0.48\columnwidth}
        \includegraphics[width=\textwidth]{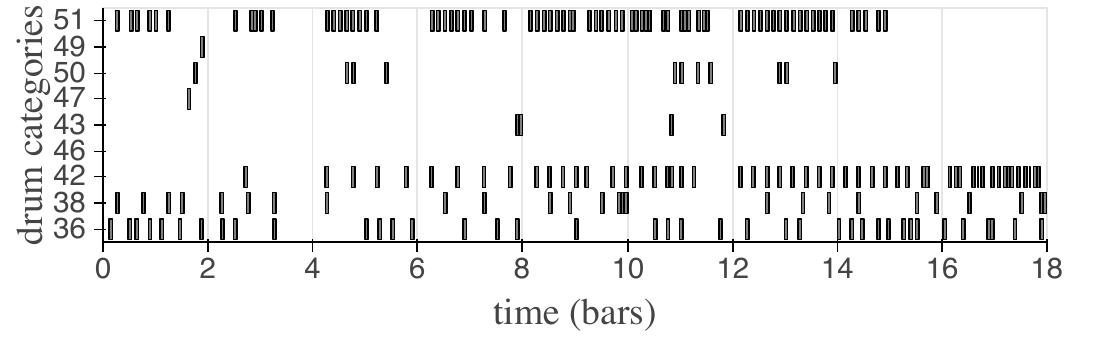}
        \caption{MusicVAE (32D)}%
    \end{subfigure}
    \\
    \begin{subfigure}[b]{0.48\columnwidth}
        \includegraphics[width=\textwidth]{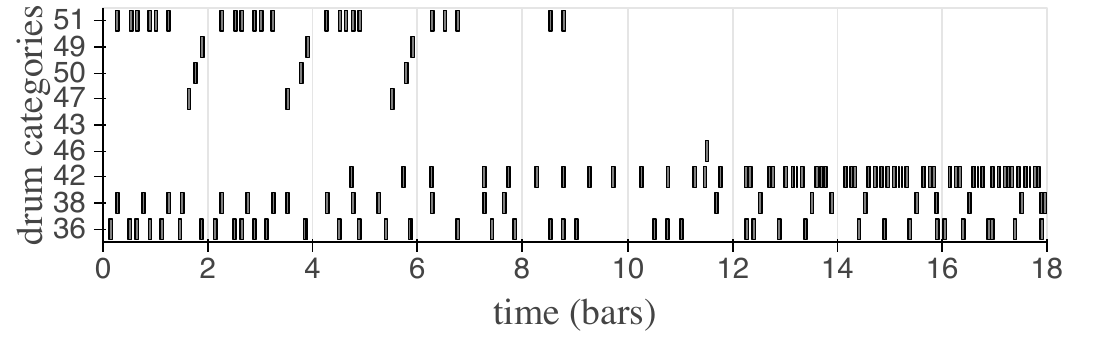}
        \caption{MusicVAE (512D)}%
    \end{subfigure}
    \caption{Drum interpolation. See \cref{fig:interp_samples} for more details.}
\end{figure}

\pagebreak

\subsection{Drum density attribute}
\label{app:drum_density}

\begin{figure}[ht!]
    \centering
    \begin{subfigure}[b]{0.48\columnwidth}
        \includegraphics[width=\textwidth]{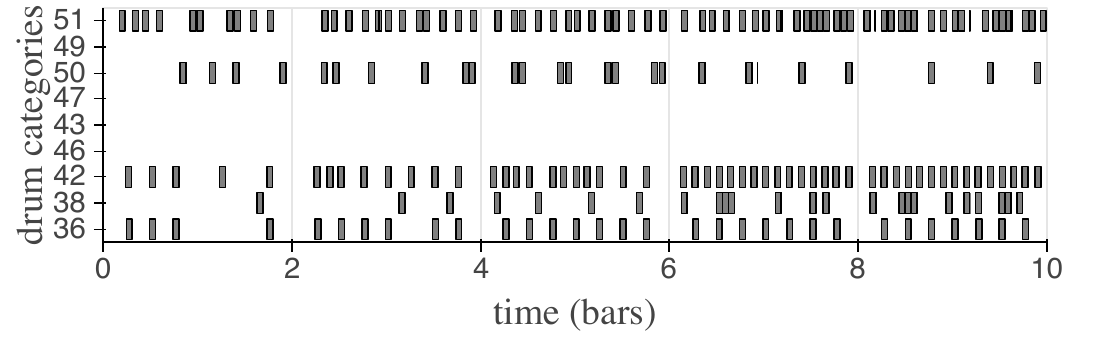}
        \caption{\ourModel}%
        \label{fig:density_musicfmvae}
    \end{subfigure}
    \\
    \begin{subfigure}[b]{0.48\columnwidth}
        \includegraphics[width=\textwidth]{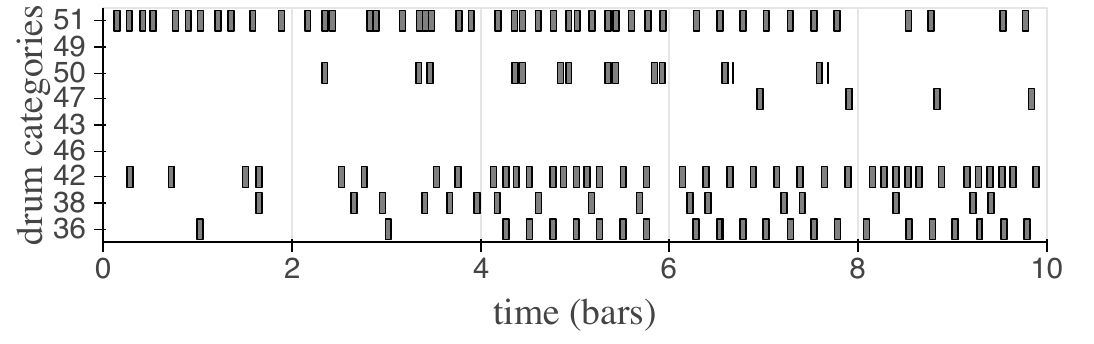}
        \caption{MusicVAE (32D)}%
        \label{fig:density_musicvae}
    \end{subfigure}
    \\
    \begin{subfigure}[b]{0.48\columnwidth}
        \includegraphics[width=\textwidth]{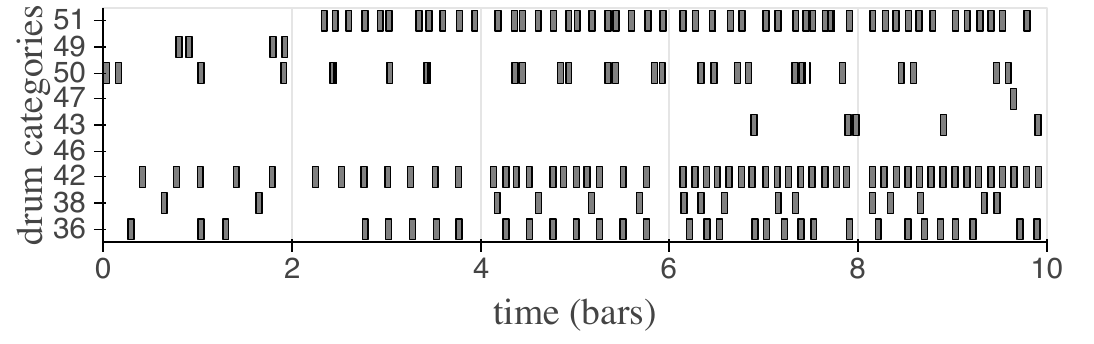}
        \caption{MusicVAE (512D)}%
        \label{fig:density_musicvae_512}
    \end{subfigure}
    \caption{Drum density attribute. See \cref{fig:density_music} for more details.}
\end{figure}

\begin{figure}[ht!]
    \centering
    \begin{subfigure}[b]{0.48\columnwidth}
        \includegraphics[width=\textwidth]{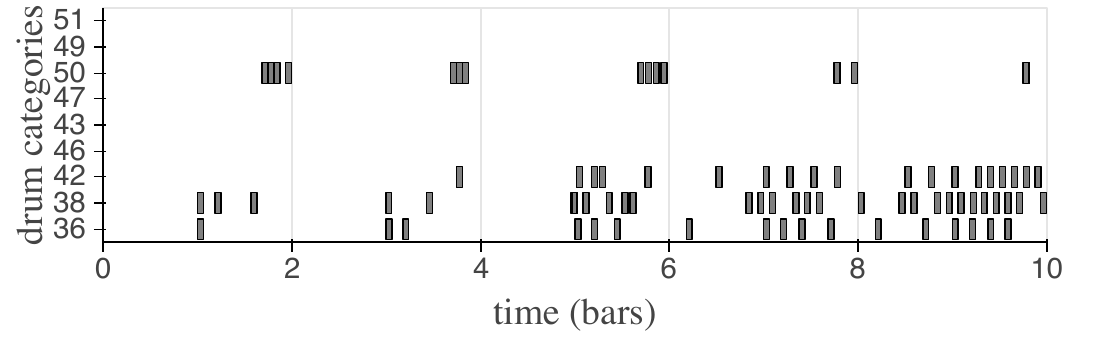}
        \caption{\ourModel}%
        \label{fig:density_musicfmvae}
    \end{subfigure}
    \\
    \begin{subfigure}[b]{0.48\columnwidth}
        \includegraphics[width=\textwidth]{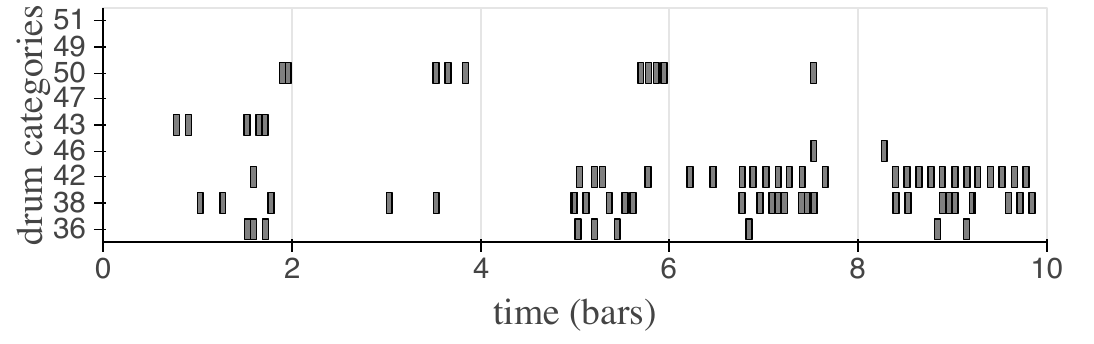}
        \caption{MusicVAE (32D)}%
        \label{fig:density_musicvae}
    \end{subfigure}
    \\
    \begin{subfigure}[b]{0.48\columnwidth}
        \includegraphics[width=\textwidth]{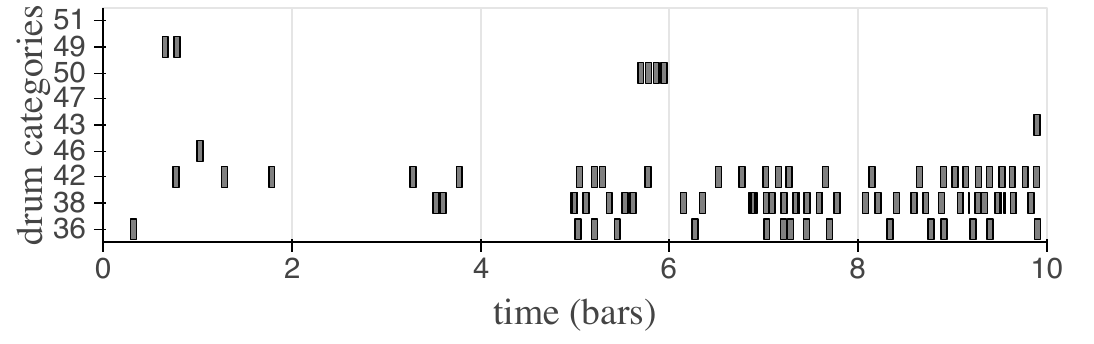}
        \caption{MusicVAE (512D)}%
        \label{fig:density_musicvae_512}
    \end{subfigure}
    \caption{Drum density attribute. See \cref{fig:density_music} for more details.}
\end{figure}

\pagebreak

\subsection{Drum velocity attribute}
\label{app:drum_velocity}

\begin{figure}[ht!]
    \centering
    \begin{subfigure}[b]{0.48\columnwidth}
        \includegraphics[width=\textwidth]{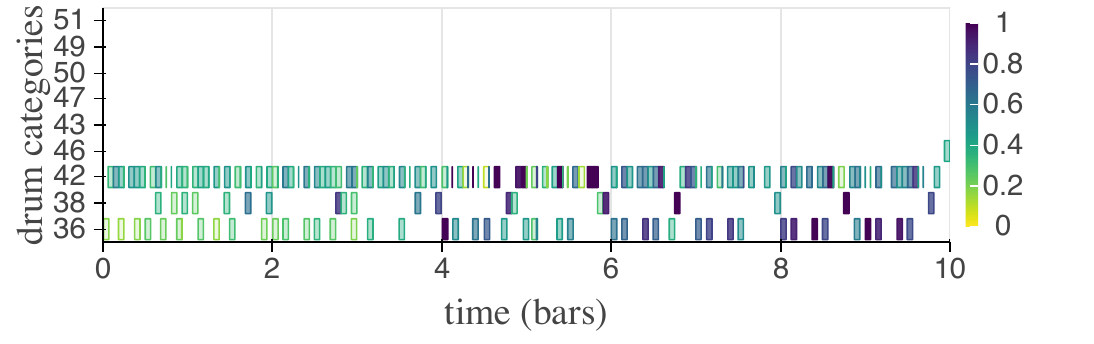}
        \caption{\ourModel}%
        \label{fig:density_musicfmvae}
    \end{subfigure}
    \\
    \begin{subfigure}[b]{0.48\columnwidth}
        \includegraphics[width=\textwidth]{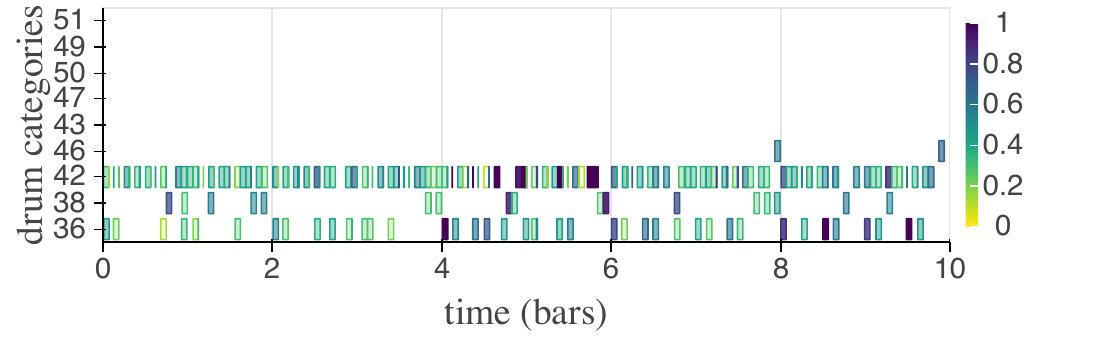}
        \caption{MusicVAE (32D)}%
        \label{fig:density_musicvae}
    \end{subfigure}
    \\
    \begin{subfigure}[b]{0.48\columnwidth}
        \includegraphics[width=\textwidth]{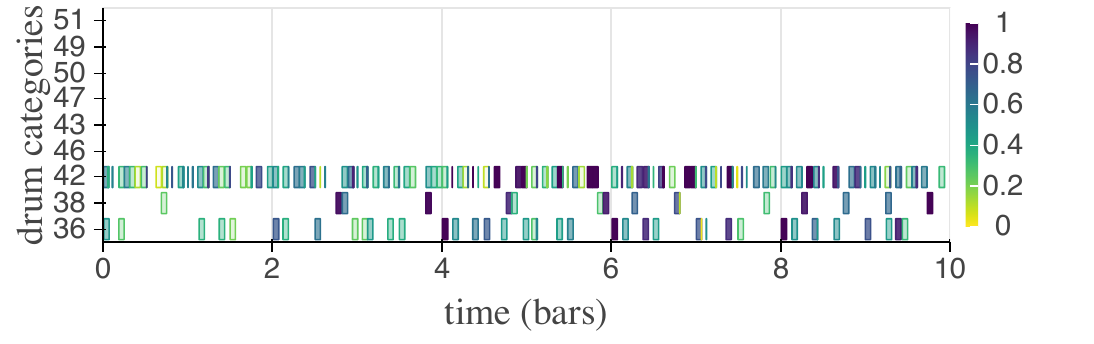}
        \caption{MusicVAE (512D)}%
        \label{fig:density_musicvae_512}
    \end{subfigure}
    \caption{Drum velocity distribute. The colour represents the velocity. See \cref{fig:density_music} for more details.}
\end{figure}

\begin{figure}[ht!]
    \centering
    \begin{subfigure}[b]{0.48\columnwidth}
        \includegraphics[width=\textwidth]{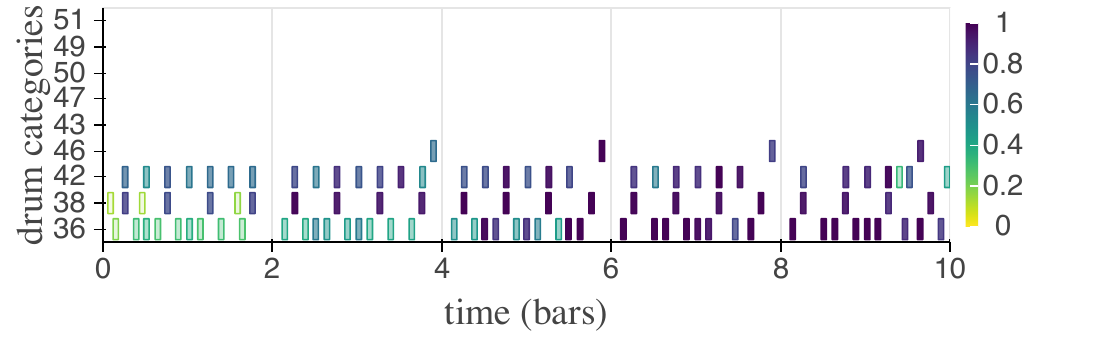}
        \caption{\ourModel}%
        \label{fig:density_musicfmvae}
    \end{subfigure}
    \\
    \begin{subfigure}[b]{0.48\columnwidth}
        \includegraphics[width=\textwidth]{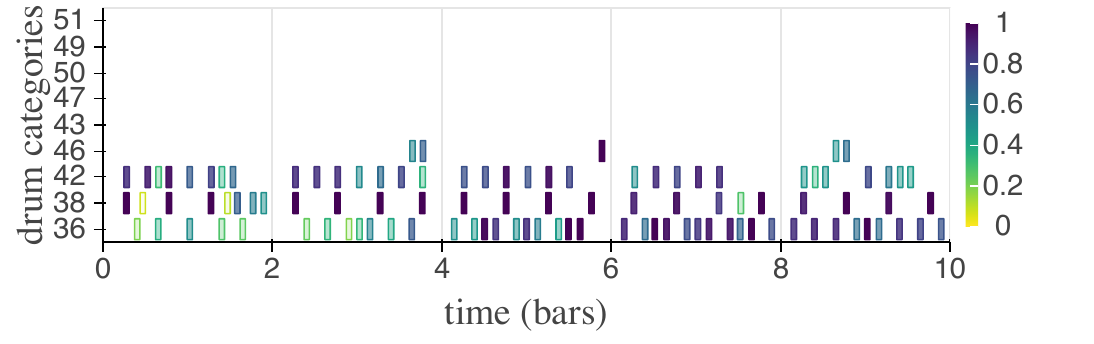}
        \caption{MusicVAE (32D)}%
        \label{fig:density_musicvae}
    \end{subfigure}
    \\
    \begin{subfigure}[b]{0.48\columnwidth}
        \includegraphics[width=\textwidth]{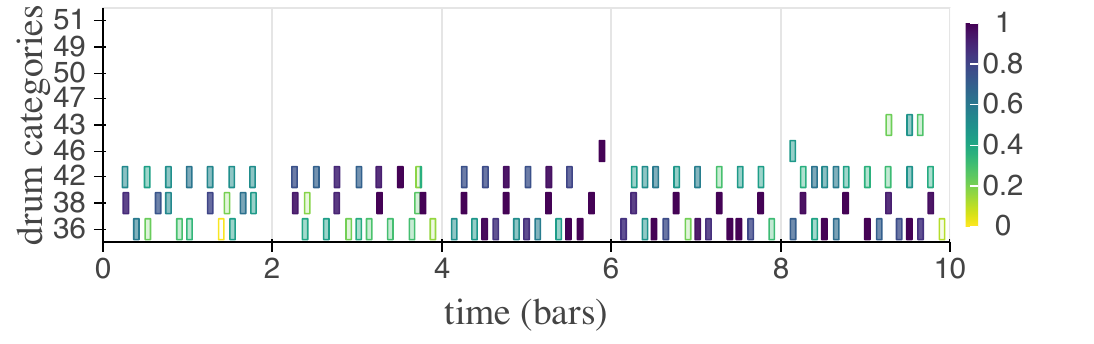}
        \caption{MusicVAE (512D)}%
        \label{fig:density_musicvae_512}
    \end{subfigure}
    \caption{Drum velocity attribute. The colour represents the velocity. See \cref{fig:density_music} for more details.}
\end{figure}

\pagebreak

\subsection{Interpolation of the piano dataset}
\label{app:interp_piano}

\begin{figure}[ht!]
    \centering
    \begin{subfigure}[b]{0.48\columnwidth}
        \includegraphics[width=\textwidth]{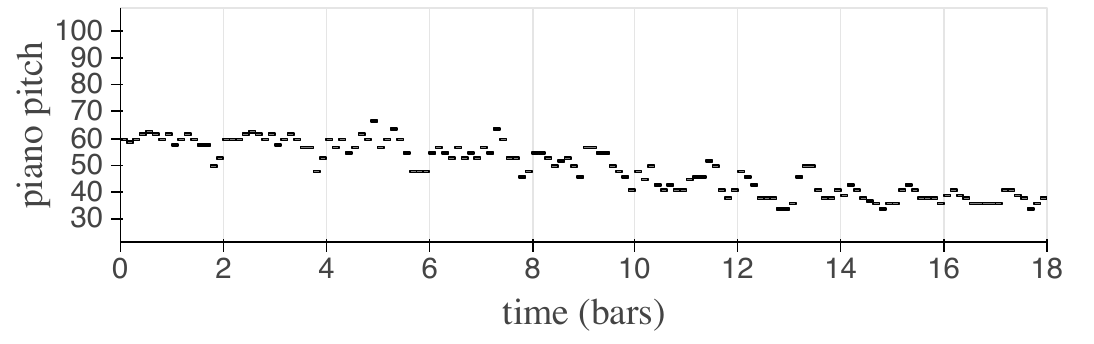}
        \caption{\ourModel}%
    \end{subfigure}
    \\
    \begin{subfigure}[b]{0.48\columnwidth}
        \includegraphics[width=\textwidth]{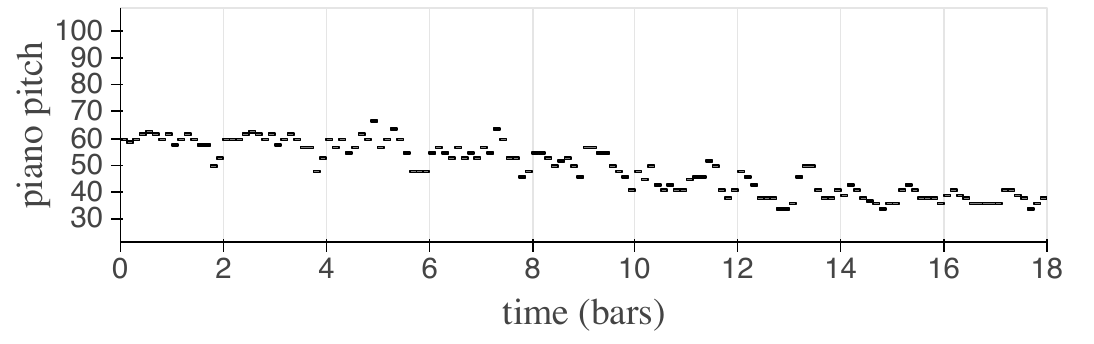}
        \caption{MusicVAE (32D)}%
    \end{subfigure}
    \\
    \begin{subfigure}[b]{0.48\columnwidth}
        \includegraphics[width=\textwidth]{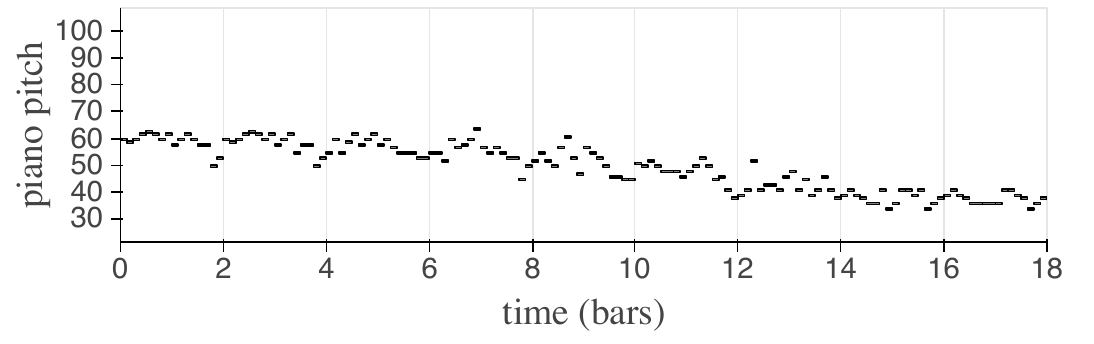}
        \caption{MusicVAE (512D)}%
    \end{subfigure}
    \caption{Piano interpolation. See \cref{fig:interp_samples} for more details.}
\end{figure}

\begin{figure}[ht!]
    \centering
    \begin{subfigure}[b]{0.48\columnwidth}
        \includegraphics[width=\textwidth]{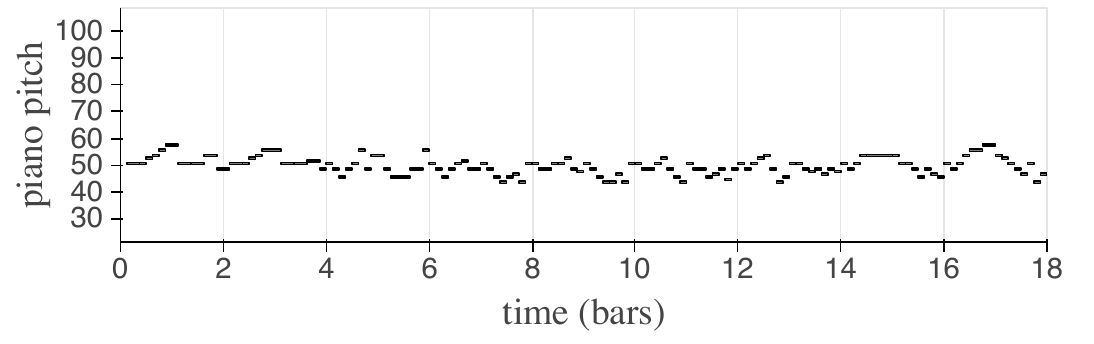}
        \caption{\ourModel}%
    \end{subfigure}
    \\
    \begin{subfigure}[b]{0.48\columnwidth}
        \includegraphics[width=\textwidth]{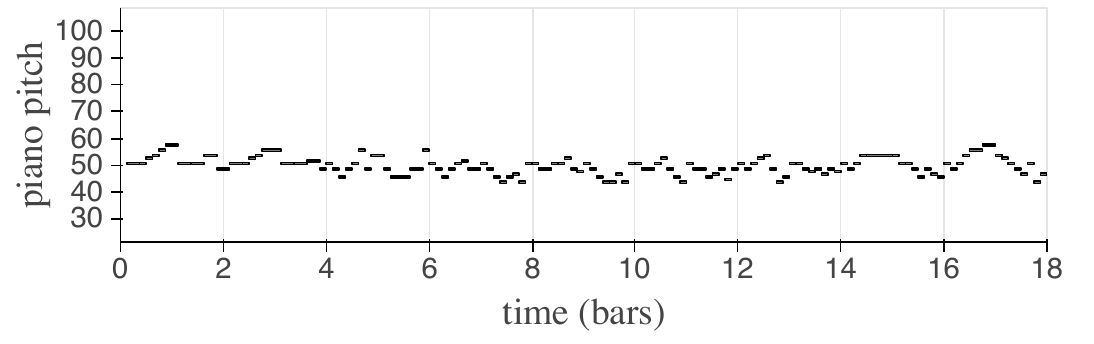}
        \caption{MusicVAE (32D)}%
    \end{subfigure}
    \\
    \begin{subfigure}[b]{0.48\columnwidth}
        \includegraphics[width=\textwidth]{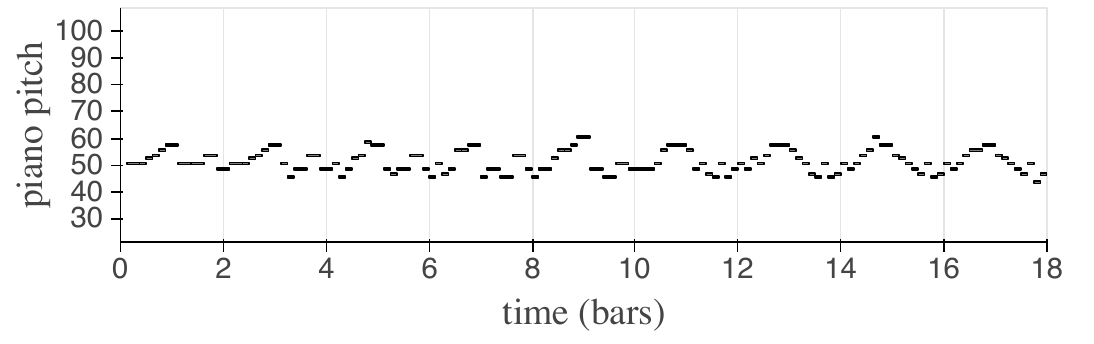}
        \caption{MusicVAE (512D)}%
    \end{subfigure}
    \caption{Piano interpolation. See \cref{fig:interp_samples} for more details.}
\end{figure}

\pagebreak

\subsection{Piano density attribute}
\label{app:piano_density}

\begin{figure}[ht!]
    \centering
    \begin{subfigure}[b]{0.48\columnwidth}
        \includegraphics[width=\textwidth]{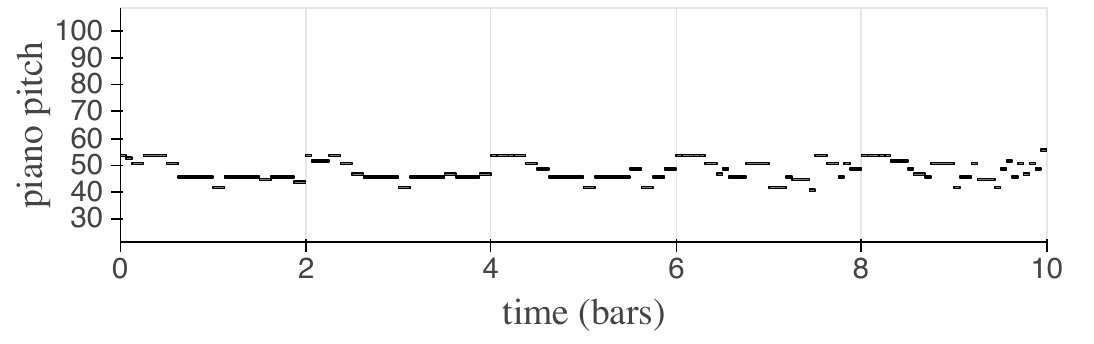}
        \caption{\ourModel}%
    \end{subfigure}
    \\
    \begin{subfigure}[b]{0.48\columnwidth}
        \includegraphics[width=\textwidth]{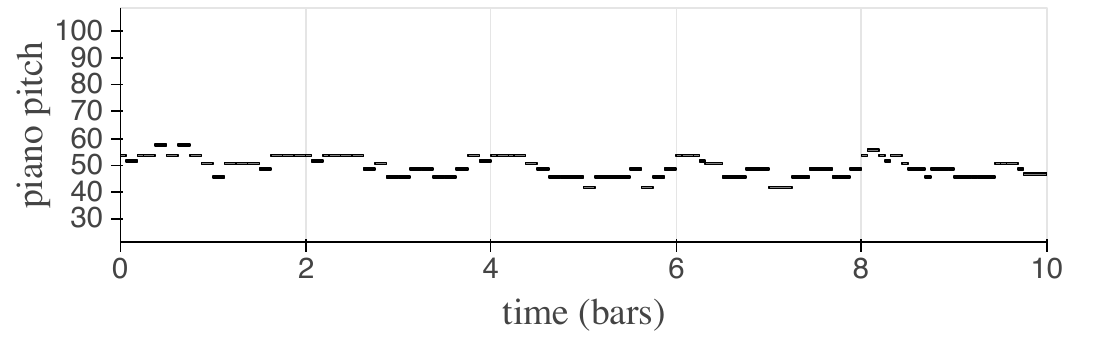}
        \caption{MusicVAE (32D)}%
    \end{subfigure}
    \\
    \begin{subfigure}[b]{0.48\columnwidth}
        \includegraphics[width=\textwidth]{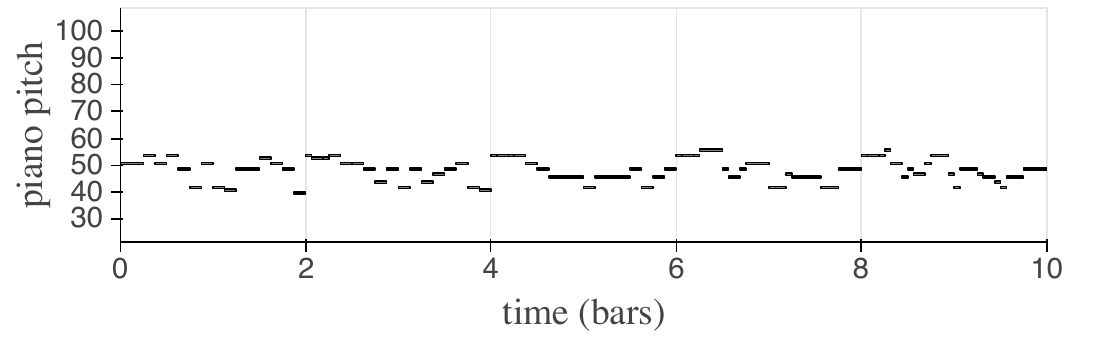}
        \caption{MusicVAE (512D)}%
    \end{subfigure}
    \caption{Piano density attribute. See \cref{fig:density_music} for more details.}
\end{figure}

\begin{figure}[ht!]
    \centering
    \begin{subfigure}[b]{0.48\columnwidth}
        \includegraphics[width=\textwidth]{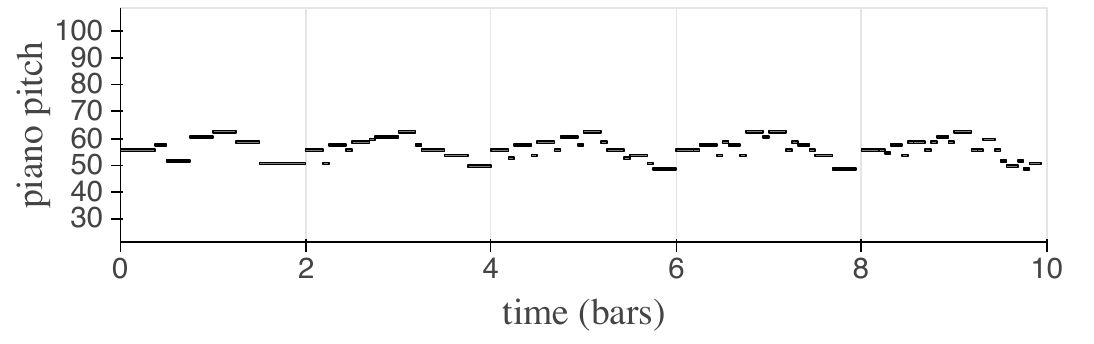}
        \caption{\ourModel}%
    \end{subfigure}
    \\
    \begin{subfigure}[b]{0.48\columnwidth}
        \includegraphics[width=\textwidth]{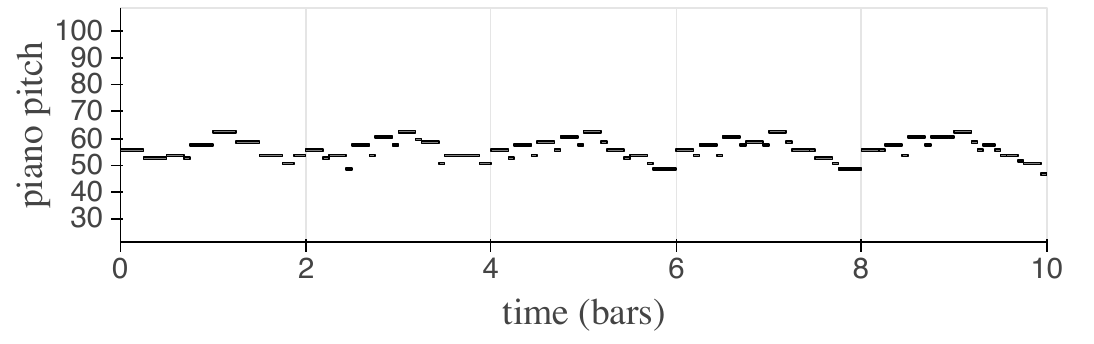}
        \caption{MusicVAE (32D)}%
    \end{subfigure}
    \\
    \begin{subfigure}[b]{0.48\columnwidth}
        \includegraphics[width=\textwidth]{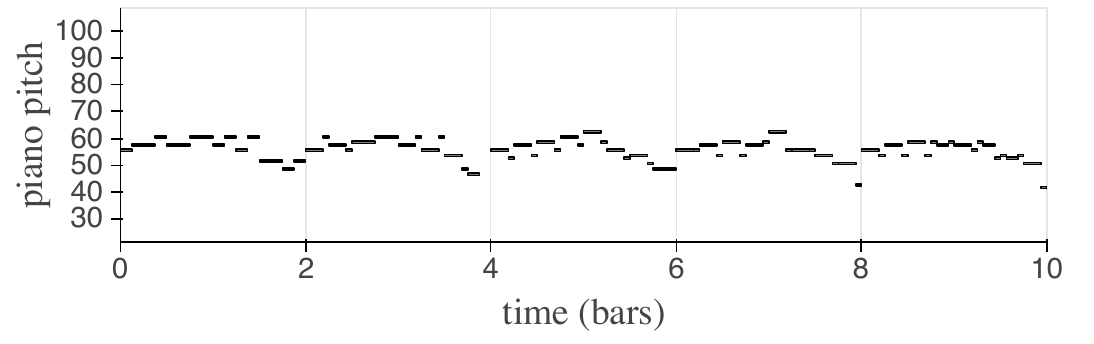}
        \caption{MusicVAE (512D)}%
    \end{subfigure}
    \caption{Piano density attribute. See \cref{fig:density_music} for more details.}
\end{figure}

\pagebreak

\subsection{Model Architectures}
\label{app:arch}
\begin{table}[!htb]
	\caption{Model architectures. FC is fully-connected layers. $\nu$ is defined in 
\protect\citep{vahiprior2019}. The models were trained on a single 16-G GPU.
}
	\label{tab:hyp}
	\vskip 0.15in
	\begin{center}
		\begin{small}
			\begin{sc}
				\begin{tabular}{llll}
					\toprule
					Dataset		& 	Optimiser		& 	 Architecture		&			 \\
					\midrule
					Drum(four bars)	& 	Adam		&	Input				&	64$\times$27	 \\
					&5$e$-4 w/ decay		&	Latents			&	32	\\
					&				&	$q_\phi(\mathbf{z}\vert\mathbf{x})$				& 	BiGRU 1024, 1024. 	\\
					&				&	$p_\theta(\mathbf{x}\vert\mathbf{z})$				&	Conductor 512, GRU 512, 512.		\\
					&				&	$q_\Phi(\mathbf{\zeta}\vert\mathbf{z})$				&	FC 256, 256, ReLU activation. 	\\
					&				&	$p_\Theta(\mathbf{z}\vert\mathbf{\zeta})$		&	FC 256, 256, ReLU activation. 	\\
					&				&	Others			& 	$\kappa$ = 0.15, $\nu$ = 1, $K$ = 16, $\eta = 2500$.\\
					\midrule					Piano		& 	Adam		&	Input				&	32$\times$90	 \\
					&5$e$-4 w/ decay		&	Latents			&	64	\\
					&				&	$q_\phi(\mathbf{z}\vert\mathbf{x})$				& 	BiGRU 1024, 1024. 	\\
					&				&	$p_\theta(\mathbf{x}\vert\mathbf{z})$				&	Conductor 512, GRU 512, 512. \\
					&				&	$q_\Phi(\mathbf{\zeta}\vert\mathbf{z})$				&	FC 256, 256, ReLU activation. 	\\
					&				&	$p_\Theta(\mathbf{z}\vert\mathbf{\zeta})$		&	FC 256, 256, ReLU activation. 	\\
					&				&	Others			& 	$\kappa$ = 0.03, $\nu$ = 1, $K$ = 8, $\eta = 2000$.\\
					\midrule					Cello		& 	Adam		&	Input				&	64$\times$90	 \\
					&5$e$-4 w/ decay		&	Latents			&	64	\\
					&				&	$q_\phi(\mathbf{z}\vert\mathbf{x})$				& 	BiGRU 1024, 1024. 	\\
					&				&	$p_\theta(\mathbf{x}\vert\mathbf{z})$				&	Conductor 512, GRU 512, 512. \\
					&				&	$q_\Phi(\mathbf{\zeta}\vert\mathbf{z})$				&	FC 256, 256, ReLU activation. 	\\
					&				&	$p_\Theta(\mathbf{z}\vert\mathbf{\zeta})$		&	FC 256, 256, ReLU activation. 	\\
					&				&	Others			& 	$\kappa$ = 0.06, $\nu$ = 1, $K$ = 8, $\eta = 2000$.\\
					\bottomrule
				\end{tabular}
			\end{sc}
		\end{small}
	\end{center}
	\vskip -0.1in
\end{table}

\subsection{Drum categories}
\label{app:drum_categories}
\begin{table}[ht!]
\vskip -0.1in
\caption{Drum categories. The mapping from the pitch to the nine drum categories. \protect\cite{GrooveVAE19} shows more details of the mapping.}
\begin{center}
\begin{footnotesize}
\begin{sc}				
\begin{tabular}{lc}
\toprule
Drum Category & Pitch \\
\midrule
Bass 	& 36 \\
Snare 	& 38 \\
High Tom 	& 50 \\
Low-Mid Tom		& 47 \\
High Floor Tom 	&43\\
Open Hi-Hat 		&46\\
Closed Hi-Hat		&42\\
Crash Cymbal 		&49\\
 Ride Cymbal 		&51\\
\bottomrule
\end{tabular}
\label{tab:drum_categories}
\end{sc}
\end{footnotesize}
\end{center}
\vskip -0.1in
\end{table}

\newpage

\end{document}